# Spin torque driven electron paramagnetic resonance of a single spin in a pentacene molecule

Stepan Kovarik,[1]†‡, Richard Schlitz[1]†§, Aishwarya Vishwakarma[1], Dominic Ruckert,[1] Pietro Gambardella[1]*, Sebastian Stepanow[1]*

**Affiliations:**

[1]Department of Materials, ETH Zurich; CH-8093 Zürich, Switzerland.

†These authors contributed equally to this work.

‡Present address: nanotech@surfaces Laboratory, Empa − Swiss Federal Laboratories for Materials Science and Technology, CH-8600 Dübendorf, Switzerland.

§Present address: Department of Physics, University of Konstanz, 78457 Konstanz, Germany.

*Corresponding author. Email: pietro.gambardella@mat.ethz.ch, sebastian.stepanow@mat.ethz.ch

**Abstract:**

Control over quantum systems is typically achieved by time-dependent electric or magnetic fields. Alternatively, electronic spins can be controlled by spin-polarized currents. Here we demonstrate coherent driving of a single spin by a radiofrequency spin-polarized current injected from the tip of a scanning tunneling microscope into an organic molecule. With the excitation of electron paramagnetic resonance, we established dynamic control of single spins by spin torque using a local electric current. In addition our work highlights the dissipative action of the spin-transfer torque, in contrast to the nondissipative action of the magnetic field, which allows for the manipulation of individual spins based on controlled decoherence.

**Introduction**

The initialization, control and readout of the electron spin underlie the processing and storing of information in both classical and quantum systems. In spintronic devices such as magnetic tunnel junctions, spin-polarized currents are routinely used to achieve electrical control of the magnetization via spin torques and readout of the magnetic state via the tunneling magnetoresistance (TMR) (*1*, *2*). Spin torques rely on the transfer of angular momentum between a spin current and a magnetic layer, mediated by the exchange interaction and scattering between conduction and localized electrons. This control has enabled the development of magnetic random-access memories (*3*, *4*) and spin torque nano-oscillators (*5–7*), which serve as tunable microwave sources (*8*), rectifiers (*9*), and neuromorphic emulators (*10*). Despite these significant achievements, the quantum mechanical nature of spin torques is still poorly understood. With limited exceptions (*11–15*), most theoretical and experimental realizations of spin torques consider the magnetic moments subjected to the torque as classical vectors obeying Landau-Lifshitz-Gilbert dynamics, which is appropriate for mesoscopic devices but not for single spins. On the other hand, methods that rely on radio-frequency (RF) magnetic fields (*16*), electric fields (*17*, *18*) and optical probes (*19*) for achieving control of single spins are well established and have widespread applications in quantum information technology, quantum sensing, and high-resolution spectroscopy (*20*, *21*). The potential of spin polarized currents in this context is much less explored. Experiments on individual spins (*22*, *23*) using scanning tunneling microscopy (STM) with magnetic tips (*24*) have revealed that the incoherent spin flips induced by a spin-polarized DC current result in a change of the spin expectation value $\langle S \rangle$ along the tip magnetization direction (*25*). This phenomenon allows for switching the orientation of atomic-scale magnets by a local electric current (*23*, *26*). Additionally, theoretical models predict the capability of spin currents to initialize individual spins into a superposition state with arbitrary components $\langle S_{x,y,z} \rangle$ and to induce electron paramagnetic resonance (EPR) (*27–29*). Harnessing spin currents to drive the initialization and precession of single spins would open new routes for the manipulation of quantum states that are complementary to methods reliant on RF electric and magnetic fields.

Here we show that the injection of an RF spin-polarized current into the singly occupied and unoccupied molecular orbitals (SOMO and SUMO) of a pentacene molecule leads to the nonequilibrium initialization of the molecular spin and resonant excitation of EPR by spin-transfer torque. Our experiments discriminate the dissipative (damping-like) and the energy-conserving (field-like) components of the spin torque. We refer to the first as the spin-transfer torque (STT) and include the second into the response of the spin to the effective magnetic field. By comparing field- and current-driven EPR, we illustrate differences in the manipulation of single spins due to the different action of the STT compared to a magnetic field. Finally, using STT-driven EPR in an STM, we demonstrate the possibility to map the spin distribution of extended orbitals in organic molecules and to achieve strong EPR signals independent of temperature due to the strong initialization of the spin states.

**Pentacene as a single spin model system**

To study the influence of spin torques at the quantum level, we focused on pentacene ($C_{22}H_{14}$) adsorbed on a thin MgO layer deposited on an Ag(001) crystal as a model quantum dot system that hosts a single spin in a molecular orbital weakly coupled to a metallic reservoir. We used STM to probe the topography and local density of states of pentacene via the differential tunneling conductance $dI/dU$ and to locally inject a spin-polarized current. The magnetic moment of pentacene on MgO originates from one electron transferred from the Ag substrate into its lowest unoccupied molecular orbital (LUMO), resulting in the formation of SOMO and SUMO states (*30*). Accordingly, the $dI/dU$ spectrum peaks at the energy of the SOMO and SUMO below and above the Fermi level, respectively (Fig. 1A). Both orbitals present the nodal structure of the pristine LUMO (insets of Fig. 1A), evidencing that the decoupling provided by the MgO layer was sufficient to preserve the electronic structure of pentacene and that the charging electron was fully localized on the molecule, in agreement with previous studies (*30–32*).

To measure the effect of a spin polarized current on the pentacene spin, we applied an RF voltage $U_{RF}$ to the tunnel junction and utilized a magnetic tip to detect the modulation of the TMR associated with the DC and RF response of the spin (*33–35*). This technique, called EPR-STM, has recently achieved unprecedented resolution to detect and manipulate spins at surfaces (*36–42*), including the coherent manipulation of one or more atomic spins (*43, 44*) (fig. S1). Previous investigations using EPR-STM have focused on spins localized at transition-metal atoms with strong electronic hybridization with the substrate, which hinders the observation of orbital specific EPR. In these systems, the EPR is excited by a time-dependent magnetic field $B_1$ originating from the adiabatic oscillation of the probed atoms in the inhomogeneous magnetic field of the STM tip, which has both exchange and dipolar components. This oscillation is a piezoelectric response to the RF electric field associated to $U_{RF}$ (*33*, *38*, *45*). In the following, we show that a second EPR excitation mechanism exists in addition to $B_1$, which is consistent with the STT driving of a single spin and largely dominates $B_1$ upon direct electron tunneling in the SOMO and SUMO.

**EPR of a single spin in a delocalized molecular orbital**

A schematic view of the EPR-STM junction is shown in Fig. 1B. We prepared a spin-polarized STM tip by picking up several Fe atoms co-deposited with pentacene on 2 monolayers of MgO (Fig. 1C). The tip magnetization has both out-of-plane ($z$) and in-plane ($x$) components owing to the magnetic anisotropy of the Fe cluster assembled on the tip. We used lock-in modulation of $U_{RF}$ superimposed to the DC bias $U_{DC}$ to measure the change $\Delta I$ of the tunneling current $I$ due to the excitation of EPR as a function of the external magnetic field $B$ applied along $z$ (*35*). The EPR spectra measured at a temperature of 4.5 K (Fig. 1D) consist of three peaks, of which the outer two shift linearly with the frequency $f$ of $U_{RF}$, in agreement with the change of Zeeman energy expected for a spin $S = 1/2$ (fig. S2). The spatial extent of the spin density was revealed by mapping the EPR signal at constant $f$ and $B$ (*46*, *47*). The maps of $I$ and $\Delta I$ in Fig. 1E, F show resonant slices of the spin density, which is distributed over the entire molecule and corresponds to the electronic density of the SOMO (see (*48*) and figs. S3–S9 for a discussion of the EPR maps in comparison to the STM topography).

The EPR spectra confirmed the paramagnetic character of pentacene. Moreover, they evidenced two features that are not expected for EPR driven by a time-dependent magnetic field $B_1$. The first is the large amplitude of the EPR lines, which is independent of $B$ (fig. S2) despite the relatively high temperature of our experiment ($k_B T \approx 0.4$ meV) compared to the Zeeman energy splitting ($E_Z = 2\mu_B B \approx 3 - 150$ μeV) (figs. S10–S12). For $B_1$-driven EPR, in the absence of spin initialization, the signal amplitude should be proportional to the difference in thermal population of the $\pm S_z$ spin states, which cannot be resolved for a single spin with $E_Z \sim f \approx 1$ GHz at 4.5 K. The second feature is the peak near $B = 0$, which has not been reported in previous EPR-STM studies and is observed also in the absence of RF excitation (*48*) (fig. S13). This feature is consistent with the change of conductance induced by a Hanle-like precession of the transverse $\langle S_x \rangle$ component injected in the SOMO in the magnetic field experienced by the molecule, as expected for a quantum dot attached to magnetic leads (*11*, *12*, *27*, *49*, *50*). The dependence of the zero field conductance on current (fig. S13), external field and tip magnetization is rather complex and will be considered in future work (*48*).

To further explore the origin of the EPR signal, we studied its dependence on the DC bias, which determines whether tunneling proceeds via the electronic states of the substrate or the SOMO and SUMO of pentacene. EPR spectra recorded in the SOMO-SUMO gap ($U_{DC} = 250$ mV) and near the SUMO ($U_{DC} = 800$ mV) exhibit a markedly different behavior as a function of the excitation frequency (Fig. 2A). The amplitude of the EPR signal at $U_{DC} = 250$ mV (yellow squares in Fig. 2B) increases linearly with $f \sim B$, consistent with the polarization of a spin $S = 1/2$, proportional to $\tanh(E_Z/2k_BT) \approx E_Z/2k_BT$ (*42, 51*). At $U_{DC} = 800$ mV instead (blue dots), the spin-polarized current through the SUMO leads to a frequency-independent EPR intensity that is much larger than the Boltzmann limit, corresponding to the spin polarization expected at mK temperature. We observed the same behavior when tunneling through the SOMO. Thus, we conclude that resonant tunneling into the occupied and unoccupied molecular orbitals efficiently polarized the molecule and that this initialization by the spin-polarized current dominated over all other relaxation channels (*48*) (fig. S14).

The strong initialization of the spin, which was further confirmed by the dependence of the tip standoff distance on bias and magnetic field (*48*) (figs. S11 and S12), can be assigned to a rotation of the spin polarization vector towards the tip magnetization (*27, 52*). If the latter has a finite *x* component, this rotation corresponds to a spin-transfer torque, leading to a superposition state $|S_x\rangle = (|S_z\rangle + |-S_z\rangle)/\sqrt{2}$. Thus, the spin is not in an eigenstate and will precess in the magnetic field. In the following, we show that synchronizing the STT with the Larmor precession of the spin in the magnetic field leads to EPR.

**Experimental signatures of spin-torque driven EPR**

We analyzed the EPR lineshape and its bias dependence to reveal telltale signatures of spin torque driving at the single-spin level. We proceeded by analogy with mesoscopic devices, where the ferromagnetic resonance driven by STT is distinguished from that driven by an oscillating magnetic field $B_1$ by the symmetry of the resonance lineshape (*6, 7, 9, 53*). In general, the EPR-STM signal can be decomposed into the sum of symmetric ($\Delta I_s$) and antisymmetric ($\Delta I_a$) Lorentzian functions

$$\Delta I = \Delta I_s \frac{1}{1+\epsilon^2} + \Delta I_a \frac{\epsilon}{1+\epsilon^2} + \Delta I_{\text{off}}, \tag{1}$$

where $\epsilon = \frac{B-B_0}{\Gamma/2}$, $B_0$ and $\Gamma$ are the resonance field and width of the EPR line, respectively, and $\Delta I_{\text{off}}$ is an offset current due to nonmagnetic rectification effects. For the $B_1$ driving established for single atoms (*34, 38, 45*), $\Delta I_s$ arises from the time-averaged change of the projection of the atom's spins on the tip magnetization, whereas $\Delta I_a$ arises from the homodyne mixing of $U_{\text{RF}}$ and time-dependent TMR caused by the rotating component of the spin (*34, 38*). The EPR spectra in Fig. 2C and the bias dependence of $\Delta I_s$ and $\Delta I_a$ in Fig. 2D show significant deviations from previous studies of single atoms in which EPR was induced by $B_1$ (*34, 38, 51*).

At low bias, we found that $\Delta I_s$ has the same sign for both positive and negative $U_{\text{DC}}$, consistent with $B_1$-driven EPR and the simultaneous inversion of the TMR and tunneling current with DC bias in pentacene (*48*). As $|U_{\text{DC}}|$ approached the energy of the SOMO and SUMO, however, both $\Delta I_s$ and $\Delta I_a$ increased strongly due to the more efficient initialization and the amplification of the TMR with the growing fraction of current flowing through the molecular orbitals. Upon further increasing $|U_{\text{DC}}|$ beyond the onset of the SOMO and SUMO resonances, the conductance increased, and the tip retracted to maintain a constant current (*48*). This process in turn led to a much less efficient $B_1$ driving, as shown by the reduction of $\Delta I_a$ already at $U_{\text{DC}}$ lower than the SOMO and SUMO peaks in the conductance spectrum (Fig. 2D). $\Delta I_s$, on the other hand, changed sign on the flanks of the SOMO and SUMO. This behavior is unexpected, as changing the sign of $\Delta I_s$ should also lead to an inversion of $\Delta I_a$ for $B_1$-driven EPR (see Methods). Moreover, the complex dependence of $\Delta I_{s,a}$ on bias cannot be ascribed to changes of the rectified current near the SOMO and SUMO (*48*). The two sign changes of $\Delta I_s$ at $U_{\text{DC}} = -350$ mV and 650 mV thus indicate the emergence of an EPR driving mechanism that only generates a symmetric signal upon tunneling through the molecular orbitals and has opposite sign relative to $\Delta I_s$ in the SOMO-SUMO gap.

We associate the symmetric lineshape observed when a spin-polarized current tunnels into the SOMO and SUMO to STT-driven EPR. This assignment agrees with theoretical models of coherent quantum transport in an STM junction under an RF bias (*27–29*) as well as with the basic properties of angular momentum transfer via a spin transfer mechanism. For a single spin in a tunnel junction, STT driving appears through the periodic variation of the initialization rate caused by the modulation of the tunneling probability and the coherent evolution of $\langle \mathbf{S} \rangle$ in a magnetic field (*28, 29*). If the STT modulation rate matches the Larmor frequency $\omega_\mathrm{L} = \gamma B$ ($\gamma$ is the electronic gyromagnetic ratio), the precession of the in-plane expectation value $\langle \mathbf{S}_{x,y} \rangle = \langle S_x \rangle (\hat{\mathbf{x}} \cos \omega_\mathrm{L} t + \hat{\mathbf{y}} \sin \omega_\mathrm{L} t)$ about $B$ is phase-locked to $U_\mathrm{RF}$, leading to EPR. The symmetric EPR lineshape observed in our experiments reflects the in-phase oscillation of $\langle S_x(t) \rangle$ averaged over many subsequent tunneling events. Conversely, for $B_1$-driving $\langle S_x(t) \rangle$ is 90° out of phase with respect to $U_\mathrm{RF}$, giving rise to the antisymmetric lineshape $\Delta I_\mathrm{a}$ in the homodyne channel.

The phase-locking between the precession of the molecular spin and the tunneling current can be put on a formal basis by using either a modified Bloch equation model that includes the STT driving as a time-dependent relaxation rate of $\langle \mathbf{S} \rangle$ towards the direction of the tunneling spin (*34, 38*) or a Lindblad master equation approach that describes the evolution of the reduced density matrix of the molecule (*48*) (fig. S15). Figure 3A summarizes the different contributions to the lineshape of the EPR signal from STT driving ($\mathrm{STT}_z, \mathrm{STT}_x$) and $B_1$ driving ($B_{1,z}, B_{1,x}$) obtained from the modified Bloch equation model. Note that the contributions $\mathrm{STT}_z$ and $B_{1,z}$ result from the DC change of the tunneling current proportional to $\langle S_z \rangle$, whereas $\mathrm{STT}_x$ and $B_{1,x}$ result from the homodyne detection of $\langle S_x(t) \rangle$. Both $B_{1,z}$ and $\mathrm{STT}_x$ yield a symmetric lineshape, but of opposite sign, supporting the interpretation of the EPR spectra in Fig. 2C.

An additional signature of STT driving was found in the power dependence of the EPR signal. Whereas for $B_1$-driving $\Delta I_\mathrm{s}$ saturates (*38, 51*) and $\Delta I_\mathrm{a}$ increases linearly with $U_\mathrm{RF}$ (see Methods and (*48*)), the homodyned STT signal $\mathrm{STT}_x$ scales as $U_\mathrm{RF}^2$, because the modulation of the tunneling probability entered twice, once for the STT rate, and once for the homodyne readout by the TMR (*9, 28*). To demonstrate the different power scaling of the two contributions to $\Delta I_\mathrm{s}$, we plot the evolution of the EPR signal with $U_\mathrm{RF}$ (see Fig. 3B,C). From low to high $U_\mathrm{RF}$, we observe a negative $\Delta I_\mathrm{s}$ at low $U_\mathrm{RF}$, as expected for $B_1$ driving at negative DC bias, followed by a nonlinear increase and inversion of sign of $\Delta I_\mathrm{s}$, as expected for STT driving at high $U_\mathrm{RF}$. Instead, $\Delta I_\mathrm{a}$ increases almost linearly with $U_\mathrm{RF}$, consistent with $B_1$ driving. Our observations thus indicate that STT is the dominant mechanism that drives EPR of singly-occupied molecular orbitals.

**Correspondence of spin torques in the quantum and classical descriptions**

In the context of spintronics, the STT driving that emerges from tunneling via the SOMO and SUMO corresponds to the damping-like component of the spin torque, which is dissipative in nature. This torque dominates in the sequential tunneling regime where the rate of spin-polarized electrons tunneling in or out of the molecule is large. The Pauli spin blockade causes a spin filter (spin pumping) effect when tunneling from (to) the molecule into (from) the magnetic tip, conditioned by the evolution of $\langle \mathbf{S}(t) \rangle$ in the magnetic field. This process leads to a strong transverse spin accumulation and a consequent rotation of $\langle \mathbf{S} \rangle$ antiparallel (parallel) to the tip polarization when averaging over many tunneling events. The $B_1$ driving associated with the exchange field of the tip, one the other hand, corresponds to the field-like spin torque. This torque, instead, dominates in the Coulomb-blockaded regime between SOMO and SUMO, in which the dwell time of a single electron in the molecule is large and the exchange field is determined by virtual electron hopping to the magnetic tip (*11*). Notably, to account for the precession of $\langle \mathbf{S}(t) \rangle$ in phase with the driving RF voltage, it is crucial to consider the coherences of the molecular spin in the singly-occupied charge state in both cases (*28*, *29*).

Pulsed EPR measurements showed that the longitudinal spin relaxation time of pentacene is $T_1 = 140 \pm 10$ ns at a current of 10 pA (*48*) (fig. S16). Therefore, the spin relaxation rate is slower than the tunneling rate in our experiments. Because of the time-averaged nature of $\langle \mathbf{S}(t) \rangle$, considering electrons that tunnel at a rate smaller or larger than the Larmor frequency does not change the conclusions on the STT driven EPR (*48*).

**Temporal evolution of $\langle \mathbf{S} \rangle$ due to magnetic field and STT**

We finally turn to the evolution of $\langle \mathbf{S}(t) \rangle$ on the Bloch sphere for a system driven by $B_1$ and compare it to the action of the STT. To that end, we modeled the time evolution of $\langle \mathbf{S} \rangle$ by solving the modified Bloch equations. Equivalent results can be obtained from the solution of the Lindblad model (*48*). For $B_1$ driving, we found the conventional Rabi behavior, whereby $\langle S_z \rangle$ oscillates between $+S_z$ and $-S_z$ at a rate proportional to $B_1$ (Fig. 4A) and $\langle S_x \rangle$ oscillates with a 90° phase shift relative to $U_{\text{RF}}$ (Fig. 4B). In contrast, a time-dependent STT yields no Rabi oscillations, showcasing the intrinsic difference of the two driving mechanisms. For STT driving, $\langle S_z \rangle$ decays at a rate proportional to the current (Fig. 4C) and $\langle S_x \rangle$ oscillates in phase with $U_{\text{RF}}$ (Fig. 4D).

To reveal the differences in the two driving regimes, we performed a Rabi experiment by applying RF pulses of varying length to the STM junction (see Refs. (*43*) and (*48*) for further details). We observed Rabi oscillations in the $B_1$-dominated regime at low $|U_{DC}|$. However, upon further increasing $|U_{DC}|$ beyond 260 mV, the Rabi frequency dropped until no oscillations could be observed in the STT-dominated regime (Fig. 4E). The Rabi rate $\Omega$ reduced with increasing $|U_{DC}|$, in line with the reduction of $B_1$ due to the retraction of the tip (*48*). Furthermore, the relaxation time $T_{2,\text{Rabi}}$ dropped quickly when $U_{DC}$ approached the SOMO (Fig. 4F), as expected from the increased tunneling current flowing through the molecule. The reduction of $\Omega$ and $T_{2,\text{Rabi}}$ entails a strong reduction of the $B_1$-driven EPR intensity at high bias, which contrasts with the drastic increase and sign change of the EPR signal reported in Fig. 2. Therefore, the Rabi measurements further support the dominant role of STT driving when tunneling into the molecular orbitals and its distinct phenomenology relative to $B_1$-driving.

**Conclusion and Outlook**

Our data provide evidence of coherent transfer of angular momentum to a single spin by a spin-polarized current. This transfer occurs through interaction with the tunneling electrons that involve entanglement with the molecular spin followed by relaxation of the many-body wavefunction to a single spin state. STT thus allows for the initialization of spins into superposition states with arbitrary spin polarization, overcoming the spatial and thermal population constraints associated with electromagnetic field excitations. Spin rotations induced by the dissipative STT are inherently non-unitary, which limits the coherent control of the spin's phase. However, the field-like spin torque contribution to $B_1$ provides a voltage-dependent means to coherently manipulate the spin's state. Future work may focus on implementing iterative schemes to achieve fault-tolerant single-spin initialization and rotation operations (*13*) and on the investigation of the nondissipative field-like spin torque component. In perspective, spin-polarized currents may be used to control the spin of paramagnetic centers in the gate oxide of field effect transistors and magnetic tunnel junctions within quantum devices (*54*), and entangle separated magnetic qubits (*55*). These capabilities provide complementary control of quantum spin excitations relative to RF electromagnetic fields.

**References and Notes:**

1. D. C. Ralph, M. D. Stiles, Spin transfer torques. *Journal of Magnetism and Magnetic Materials* **320**, 1190–1216 (2008).
2. A. Manchon, J. Železný, I. M. Miron, T. Jungwirth, J. Sinova, A. Thiaville, K. Garello, P. Gambardella, Current-induced spin-orbit torques in ferromagnetic and antiferromagnetic systems. *Reviews of Modern Physics* **91** (2019).
3. J. Z. Sun, Spin-transfer torque switched magnetic tunnel junction for memory technologies. *Journal of Magnetism and Magnetic Materials* **559**, 169479 (2022).
4. V. Krizakova, M. Perumkunnil, S. Couet, P. Gambardella, K. Garello, Spin-orbit torque switching of magnetic tunnel junctions for memory applications. *Journal of Magnetism and Magnetic Materials* **562**, 169692 (2022).
5. M. Tsoi, A. G. M. Jansen, J. Bass, W.-C. Chiang, V. Tsoi, P. Wyder, Generation and detection of phase-coherent current-driven magnons in magnetic multilayers. *Nature* **406**, 46–48 (2000).
6. J. C. Sankey, Y. T. Cui, J. Z. Sun, J. C. Slonczewski, R. A. Buhrman, D. C. Ralph, Measurement of the spin-transfer-torque vector in magnetic tunnel junctions. *Nature Physics* **4**, 67–71 (2008).
7. L. Liu, T. Moriyama, D. C. Ralph, R. A. Buhrman, Spin-torque ferromagnetic resonance induced by the spin Hall effect. *Physical Review Letters* **106**, 1–4 (2011).
8. T. Chen, R. K. Dumas, A. Eklund, P. K. Muduli, A. Houshang, A. A. Awad, P. Durrenfeld, B. G. Malm, A. Rusu, J. Akerman, Spin-Torque and Spin-Hall Nano-Oscillators. *Proc. IEEE* **104**, 1919–1945 (2016).
9. A. A. Tulapurkar, Y. Suzuki, A. Fukushima, H. Kubota, H. Maehara, K. Tsunekawa, D. D. Djayaprawira, N. Watanabe, S. Yuasa, Spin-torque diode effect in magnetic tunnel junctions. *Nature* **438**, 339–342 (2005).
10. J. Torrejon, M. Riou, F. A. Araujo, S. Tsunegi, G. Khalsa, D. Querlioz, P. Bortolotti, V. Cros, K. Yakushiji, A. Fukushima, H. Kubota, S. Yuasa, M. D. Stiles, J. Grollier, Neuromorphic computing with nanoscale spintronic oscillators. *Nature* **547**, 428–431 (2017).
11. M. Braun, J. König, J. Martinek, Theory of transport through quantum-dot spin valves in the weak-coupling regime. *Physical Review B* **70**, 195345 (2004).
12. M. Braun, J. König, J. Martinek, Hanle effect in transport through quantum dots coupled to ferromagnetic leads. *Europhysics Letters* **72**, 294–300 (2005).
13. B. Sutton, S. Datta, Manipulating quantum information with spin torque. *Scientific Reports 2015 5:1* **5**, 1–12 (2015).
14. A. Zholud, R. Freeman, R. Cao, A. Srivastava, S. Urazhdin, Spin Transfer due to Quantum Magnetization Fluctuations. *Phys. Rev. Lett.* **119**, 257201 (2017).
15. M. D. Petrović, P. Mondal, A. E. Feiguin, P. Plecháč, B. K. Nikolić, Spintronics Meets Density Matrix Renormalization Group: Quantum Spin-Torque-Driven Nonclassical Magnetization Reversal and Dynamical Buildup of Long-Range Entanglement. *Phys. Rev. X* **11**, 021062 (2021).
16. J. J. Pla, K. Y. Tan, J. P. Dehollain, W. H. Lim, J. J. L. Morton, D. N. Jamieson, A. S. Dzurak, A. Morello, A single-atom electron spin qubit in silicon. *Nature* **489**, 541–545 (2012).
17. K. C. Nowack, F. H. L. Koppens, Yu. V. Nazarov, L. M. K. Vandersypen, Coherent control of a single electron spin with electric fields. *Science* **318**, 1430–1433 (2007).

18. A. Laucht, J. T. Muhonen, F. A. Mohiyaddin, R. Kalra, J. P. Dehollain, S. Freer, F. E. Hudson, M. Veldhorst, R. Rahman, G. Klimeck, K. M. Itoh, D. N. Jamieson, J. C. McCallum, A. S. Dzurak, A. Morello, Electrically controlling single-spin qubits in a continuous microwave field. *Science Advances* **1**, 1–6 (2015).
19. D. D. Awschalom, R. Hanson, J. Wrachtrup, B. B. Zhou, Quantum technologies with optically interfaced solid-state spins. *Nature Photon* **12**, 516–527 (2018).
20. A. J. Heinrich, W. D. Oliver, L. M. K. Vandersypen, A. Ardavan, R. Sessoli, D. Loss, A. B. Jayich, J. Fernandez-Rossier, A. Laucht, A. Morello, Quantum-coherent nanoscience. *Nature Nanotechnology 2021 16:12* **16**, 1318–1329 (2021).
21. E. Moreno-Pineda, W. Wernsdorfer, Measuring molecular magnets for quantum technologies. *Nature Reviews Physics* **3**, 645–659 (2021).
22. S. Loth, K. von Bergmann, M. Ternes, A. F. Otte, C. P. Lutz, A. J. Heinrich, Controlling the state of quantum spins with electric currents. *Nature Physics* **6**, 340–344 (2010).
23. A. A. Khajetoorians, B. Baxevanis, C. Hübner, T. Schlenk, S. Krause, T. O. Wehling, S. Lounis, A. Lichtenstein, D. Pfannkuche, J. Wiebe, R. Wiesendanger, Current-Driven Spin Dynamics of Artificially Constructed Quantum Magnets. *Science* **339**, 55–59 (2013).
24. R. Wiesendanger, Spin mapping at the nanoscale and atomic scale. *Reviews of Modern Physics* **81**, 1495–1550 (2009).
25. F. Delgado, J. J. Palacios, J. Fernández-Rossier, Spin-Transfer Torque on a Single Magnetic Adatom. *Physical Review Letters* **104**, 026601 (2010).
26. S. Loth, S. Baumann, C. P. Lutz, D. M. Eigler, A. J. Heinrich, Bistability in atomic-scale antiferromagnets. *Science* **335**, 196–199 (2012).
27. A. M. Shakirov, Y. E. Shchadilova, A. N. Rubtsov, P. Ribeiro, Role of coherence in transport through engineered atomic spin devices. *Physical Review B* **94**, 224425 (2016).
28. A. M. Shakirov, A. N. Rubtsov, P. Ribeiro, Spin transfer torque induced paramagnetic resonance. *Physical Review B* **99**, 054434 (2019).
29. J. Reina-Gálvez, C. Wolf, N. Lorente, Many-body nonequilibrium effects in all-electric electron spin resonance. *Phys. Rev. B* **107**, 235404 (2023).
30. M. Hollerer, D. Lüftner, P. Hurdax, T. Ules, S. Soubatch, F. S. Tautz, G. Koller, P. Puschnig, M. Sterrer, M. G. Ramsey, Charge Transfer and Orbital Level Alignment at Inorganic/Organic Interfaces: The Role of Dielectric Interlayers. *ACS Nano* **11**, 6252–6260 (2017).
31. J. Repp, G. Meyer, S. Paavilainen, F. E. Olsson, M. Persson, Imaging Bond Formation Between a Gold Atom and Pentacene on an Insulating Surface. *Science* **312**, 1196–1199 (2006).
32. J. Repp, G. Meyer, S. M. Stojković, A. Gourdon, C. Joachim, Molecules on Insulating Films: Scanning-Tunneling Microscopy Imaging of Individual Molecular Orbitals. *Physical Review Letters* **94**, 026803 (2005).
33. S. Baumann, W. Paul, T. Choi, C. P. Lutz, A. Ardavan, A. J. Heinrich, Electron paramagnetic resonance of individual atoms on a surface. *Science* **350**, 417–420 (2015).
34. Y. Bae, K. Yang, P. Willke, T. Choi, A. J. Heinrich, C. P. Lutz, Enhanced quantum coherence in exchange coupled spins via singlet-triplet transitions. *Science Advances* **4**, eaau4159 (2018).
35. T. S. Seifert, S. Kovarik, C. Nistor, L. Persichetti, S. Stepanow, P. Gambardella, Single-atom electron paramagnetic resonance in a scanning tunneling microscope driven by a radio-frequency antenna at 4 K. *Physical Review Research* **2**, 013032 (2020).
36. T. Choi, W. Paul, S. Rolf-Pissarczyk, A. J. MacDonald, F. D. Natterer, K. Yang, P. Willke,

C. P. Lutz, A. J. Heinrich, Atomic-scale sensing of the magnetic dipolar field from single atoms. *Nature Nanotechnology* **12**, 420–424 (2017).
37. P. Willke, Y. Bae, K. Yang, J. L. Lado, A. Ferrón, T. Choi, A. Ardavan, J. Fernández-Rossier, A. J. Heinrich, C. P. Lutz, Hyperfine interaction of individual atoms on a surface. *Science* **362**, 336–339 (2018).
38. T. S. Seifert, S. Kovarik, D. M. Juraschek, N. A. Spaldin, P. Gambardella, S. Stepanow, Longitudinal and transverse electron paramagnetic resonance in a scanning tunneling microscope. *Science Advances* **6**, eabc5511 (2020).
39. M. Steinbrecher, W. M. J. Van Weerdenburg, E. F. Walraven, N. P. E. Van Mullekom, J. W. Gerritsen, F. D. Natterer, D. I. Badrtdinov, A. N. Rudenko, V. V. Mazurenko, M. I. Katsnelson, A. Van Der Avoird, G. C. Groenenboom, A. A. Khajetoorians, Quantifying the interplay between fine structure and geometry of an individual molecule on a surface. *Physical Review B* **103**, 155405 (2021).
40. T. S. Seifert, S. Kovarik, P. Gambardella, S. Stepanow, Accurate measurement of atomic magnetic moments by minimizing the tip magnetic field in STM-based electron paramagnetic resonance. *Physical Review Research* **3**, 043185 (2021).
41. L. M. Veldman, L. Farinacci, R. Rejali, R. Broekhoven, J. Gobeil, D. Coffey, M. Ternes, A. F. Otte, Free coherent evolution of a coupled atomic spin system initialized by electron scattering. *Science* **372**, 964–968 (2021).
42. X. Zhang, C. Wolf, Y. Wang, H. Aubin, T. Bilgeri, P. Willke, A. J. Heinrich, T. Choi, Electron spin resonance of single iron phthalocyanine molecules and role of their non-localized spins in magnetic interactions. *Nature Chemistry* **14**, 59–65 (2022).
43. K. Yang, W. Paul, S.-H. H. Phark, P. Willke, Y. Bae, T. Choi, T. Esat, A. Ardavan, A. J. Heinrich, C. P. Lutz, Coherent spin manipulation of individual atoms on a surface. *Science* **366**, 509–512 (2019).
44. P. Willke, T. Bilgeri, X. Zhang, Y. Wang, C. Wolf, H. Aubin, A. Heinrich, T. Choi, Coherent Spin Control of Single Molecules on a Surface. *ACS Nano* **15**, 17959–17965 (2021).
45. J. L. Lado, A. Ferrón, J. Fernández-Rossier, Exchange mechanism for electron paramagnetic resonance of individual adatoms. *Physical Review B* **96**, 205420 (2017).
46. P. Willke, K. Yang, Y. Bae, A. J. Heinrich, C. P. Lutz, Magnetic resonance imaging of single atoms on a surface. *Nature Physics* **15**, 1005–1010 (2019).
47. S. Kovarik, R. Robles, R. Schlitz, T. S. Seifert, N. Lorente, P. Gambardella, S. Stepanow, Electron Paramagnetic Resonance of Alkali Metal Atoms and Dimers on Ultrathin MgO. *Nano Letters* **22**, 4176–4181 (2022).
48. See the supplementary materials.
49. A. D. Crisan, S. Datta, J. J. Viennot, M. R. Delbecq, A. Cottet, T. Kontos, Harnessing spin precession with dissipation. *Nat Commun* **7**, 10451 (2016).
50. P. Busz, D. Tomaszewski, J. Barnaś, J. Martinek, Hanle effect in transport through single atoms in spin-polarized STM. *Journal of Magnetism and Magnetic Materials* **588**, 171465 (2023).
51. P. Willke, W. Paul, F. D. Natterer, K. Yang, Y. Bae, T. Choi, J. F. Rossier, A. J. Heinrich, C. P. Lutz, Probing quantum coherence in single-atom electron spin resonance. *Science Advances* **4**, eaaq1543 (2018).
52. P. Mondal, U. Bajpai, M. D. Petrović, P. Plecháč, B. K. Nikolić, Quantum spin transfer torque induced nonclassical magnetization dynamics and electron-magnetization entanglement. *Phys. Rev. B* **99**, 094431 (2019).

53. A. A. Kovalev, G. E. W. Bauer, A. Brataas, Current-driven ferromagnetic resonance, mechanical torques, and rotary motion in magnetic nanostructures. *Physical Review B* **75**, 014430 (2007).
54. M. Lamblin, B. Chowrira, V. Da Costa, B. Vileno, L. Joly, S. Boukari, W. Weber, R. Bernard, B. Gobaut, M. Hehn, D. Lacour, M. Bowen, Encoding information onto the charge and spin state of a paramagnetic atom using MgO tunnelling spintronics. arXiv arXiv:2308.16592 [Preprint] (2023). http://arxiv.org/abs/2308.16592.
55. A. Suresh, R. D. Soares, P. Mondal, J. P. S. Pires, J. M. V. P. Lopes, A. Ferreira, A. E. Feiguin, P. Plecháč, B. K. Nikolić, Electron-mediated entanglement of two distant macroscopic ferromagnets within a nonequilibrium spintronic device. *Phys. Rev. A* **109**, 022414 (2024).
56. S. Loth, C. P. Lutz, A. J. Heinrich, Spin-polarized spin excitation spectroscopy. *New Journal of Physics* **12**, 125021 (2010).
57. Data for: S. Kovarik, R. Schlitz, A. Vishwakarma, D. Ruckert, P. Gambardella, S. Stepanow, Spin torque–driven electron paramagnetic resonance of a single spin in a pentacene molecule, ETH Zurich (2024); https://doi.org/10.3929/ethz-b-000669567.

**Acknowledgments:** We thank B. K. Nikolić and J. Reina-Gálvez for helpful discussions.

**Funding:**

Swiss National Science Foundation (SNSF), project No. CRSII5_205987;

ETH Research Grant ETH-45 20-2.

**Author contributions:**

Conceptualization: SK, RS, PG, SS; Funding acquisition: PG, SS; Investigation: SK, RS, AV, DR; Methodology: SK, RS, PG, SS; Project administration: PG, SS; Software: SK, RS, DR; Supervision: PG, SS; Visualization: SK, RS; Data curation: SK, RS; Writing – original draft: SK, RS, PG, SS; Writing – review & editing: SK, RS, AV, DR, PG, SS.

**Competing interests:** Authors declare that they have no competing interests.

**Data and materials availability:** All data needed to evaluate the conclusions in the paper are present in the paper or the Supplementary Materials and are available in the ETH Research Collection database (*57*).

**Supplementary Materials**

Materials and Methods; Supplementary Text; Figs. S1 to S21; References (*58–59*)

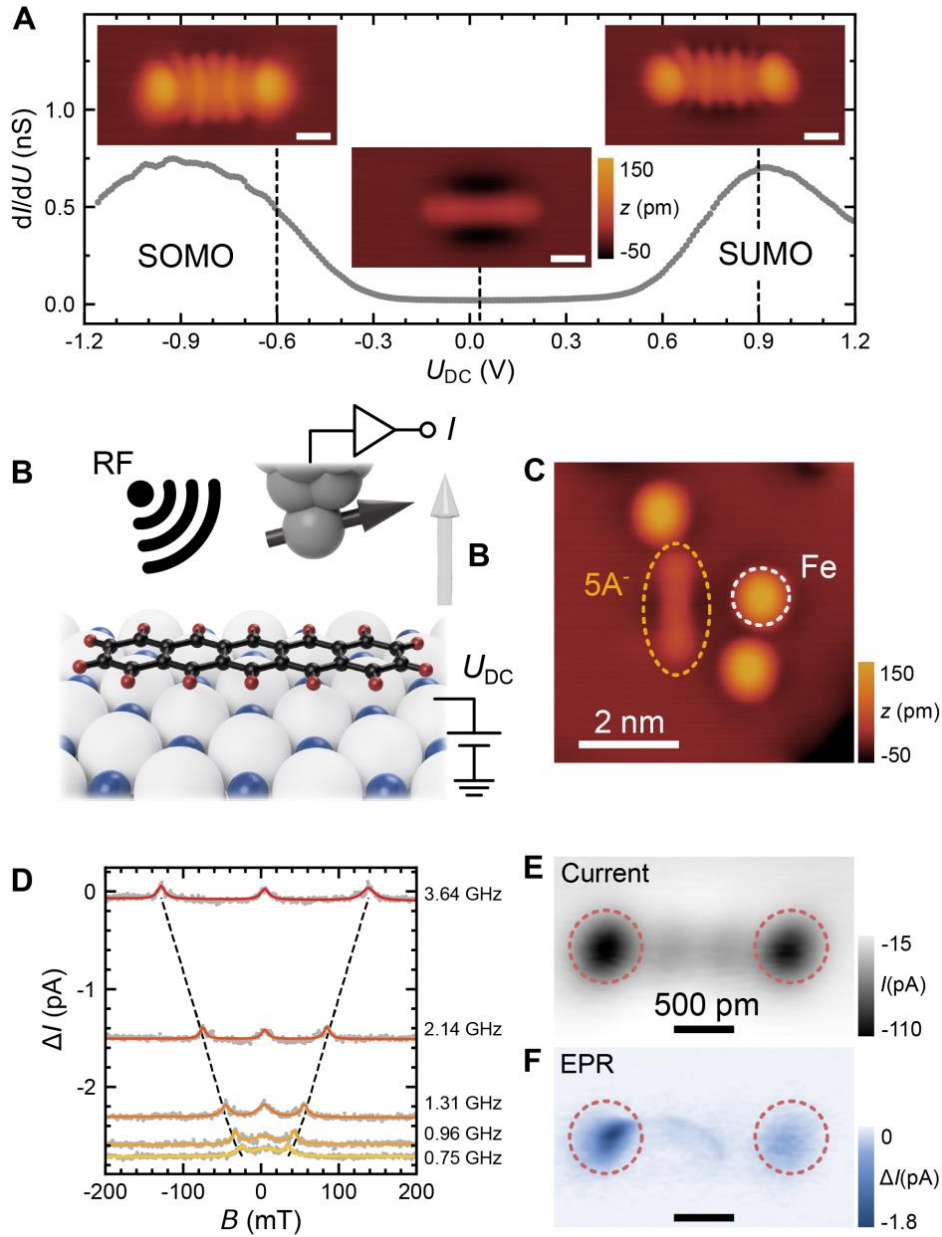

**Fig. 1. Scanning tunneling spectroscopy and EPR of individual pentacene molecules.** (**A**) Differential conductance $dI/dU$ taken at constant height (feedback opened at $I = 500$ pA and $U_{DC} = -1.2$ V). Owing to the unpaired electron in the SOMO, pentacene on MgO is an organic radical with spin $S = 1/2$. Note that, due to the vanishing density of states of the molecule near the Fermi level (56), we did not observe the usual spin-flip excitation steps near the Fermi energy in the tunneling spectra. The topography maps in the insets were taken at $I = 30$ pA and $U_{DC}$ corresponding to the dashed lines. The scale bars are 500 pm. (**B**) Sketch of the EPR-STM junction. The magnetic tip is shown on top of a pentacene molecule on MgO/Ag(100). The DC bias voltage

$U_{DC}$ was applied to the sample and an RF voltage $U_{RF}$ was capacitively coupled to the tunneling junction via an antenna. (**C**) Constant current topography of a charged pentacene molecule surrounded by three Fe atoms on 2 monolayers of MgO ($I = 30$ pA and $U_{DC} = -350$ mV). (**D**) EPR spectra as a function of $B$ acquired at different frequency ($I = 50$ pA, $U_{DC} = -520$ mV, $U_{RF} \approx 90$ mV). The dashed lines show the shift of the resonance field $B_0$ with frequency. (**E**) Current map taken at constant height (feedback opened at $I = 150$ pA and $U_{DC} = -450$ mV on the molecular lobe). (**F**) Simultaneous map of the EPR signal acquired with frequency modulation at $B = 1.28$ T and $f = 36.22$ GHz.

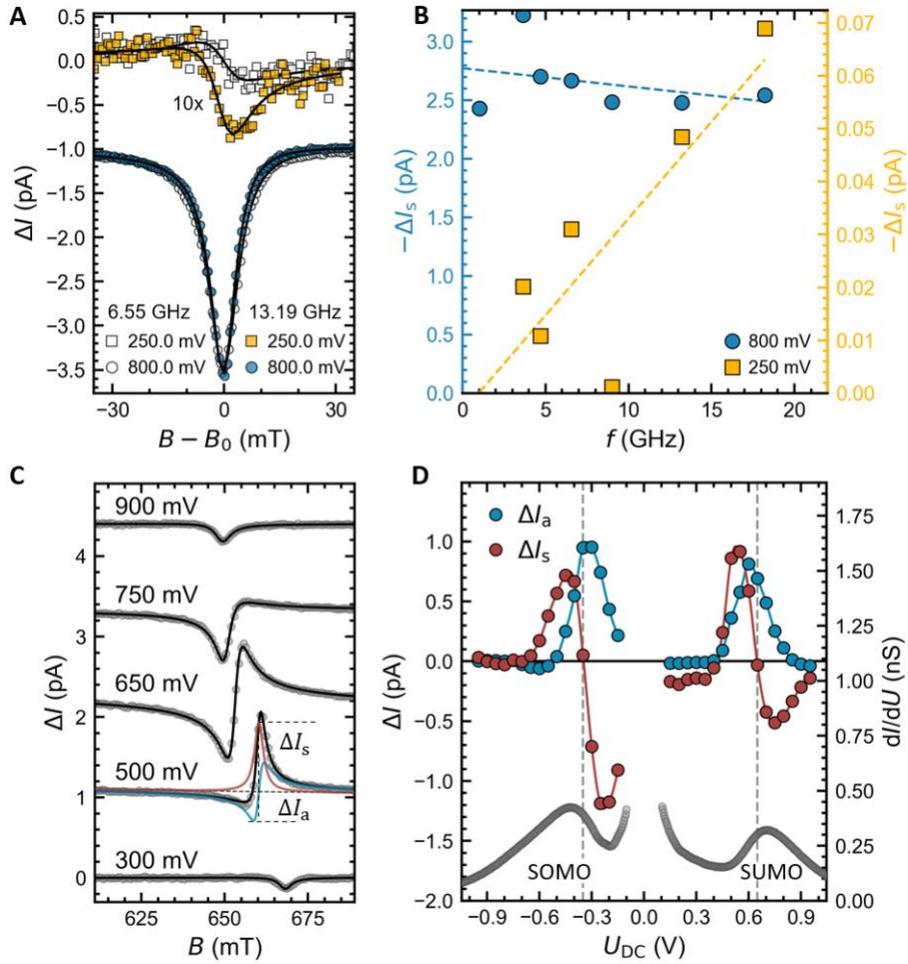

**Fig. 2. Spin-transfer torque initialization and driving of a molecular spin.** (**A**) EPR spectra as a function of $B$ acquired at $U_{DC} = 250$ mV (squares) and 800 mV (dots) at $f = 6.55$ GHz (open symbols) and 13.19 GHz (filled symbols) with $I = 100$ pA, $U_{RF} \approx 130$ mV. (**B**) Symmetric peak amplitude $\Delta I_s$ as a function of $f$ (figs. S2 and S18). The dashed lines are linear fits. (**C**) EPR spectra as a function of $B$ acquired at different $U_{DC}$ ($I = 50$ pA, $U_{RF} \approx 90$ mV, $f = 18.22$ GHz). The dashed lines correspond to the symmetric and antisymmetric components of the EPR lineshape with their respective amplitudes $\Delta I_s$ and $\Delta I_a$. (**D**) $\Delta I_s$ and $\Delta I_a$ as a function of $U_{DC}$ (figs. S17 and S19). The gray points represent the differential conductance of the molecule. The vertical dashed lines indicate the sign inversion of $\Delta I_s$ as STT driving overcomes $B_1$ driving.

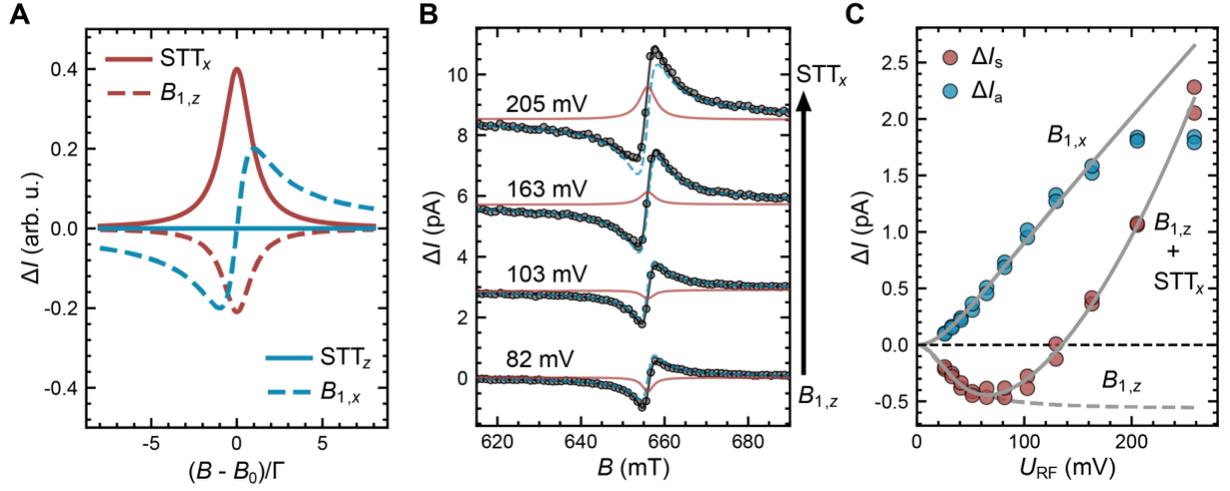

**Fig. 3. Power dependence of the EPR signal.** (**A**) Calculated lineshape of the different contributions to the EPR signal for STT and $B_1$ driving(*48*). (**B**) EPR spectra for different $U_{RF}$ acquired with $I = 50$ pA and $U_{DC} = -350$ mV. The solid lines show the symmetric (red) and antisymmetric (dashed blue) component of the fit according to Eq. (1) (black). (**C**) $\Delta I_s$ and $\Delta I_a$ as a function of $U_{RF}$. The lines are fits to the data for the different driving mechanisms (see Methods and (*48*)) The deviation of $\Delta I_a$ from the expected linear trend at large $U_{RF}$ is ascribed to the bias-dependent junction conductance around $U_{DC} = -350$ mV (see Fig. 2D and fig. S20).

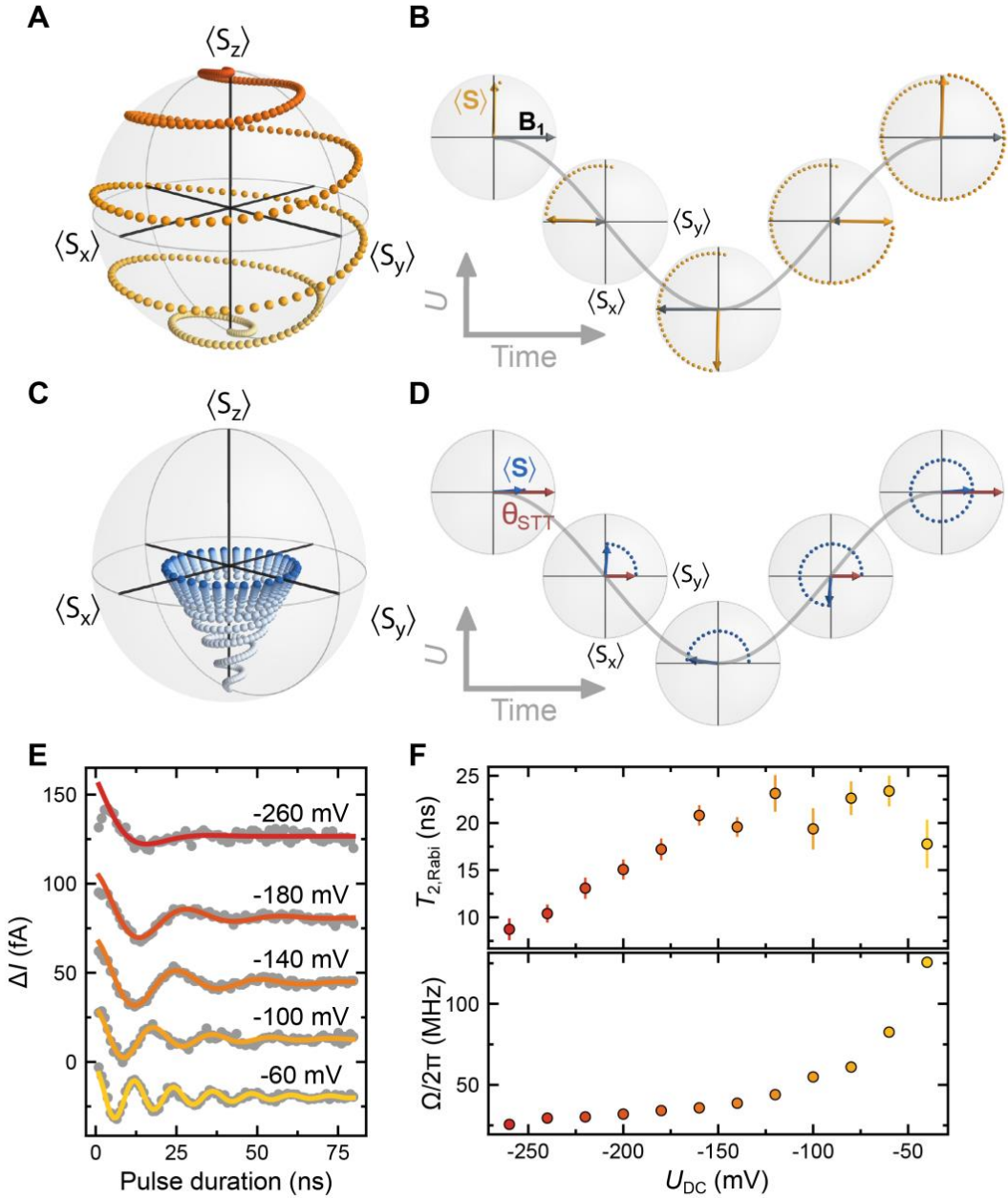

**Fig. 4. Evolution of $\langle S_z \rangle$ and $\langle S_{x,y} \rangle$ on resonance.** (**A**) Evolution of the spin expectation value for $B_1$ driven EPR during the initial half Rabi cycle. (**B**) The five circular diagrams depict the time evolution of $\langle S_{x,y} \rangle$ (yellow) and $B_1$ (gray) for $B_1$-driven EPR. The diagrams are superposed to the time evolution of the RF voltage. (**C**) Evolution of the spin expectation value for STT-driven EPR. (**D**) The five circular diagrams depict the time evolution of $\langle S_{x,y} \rangle$ (blue) and the RF STT ($\theta_{STT}$, red) superposed to $U_{RF}(t)$. (**E**) Rabi oscillations measured at $I = 13$ pA and $U_{RF} = 240$ mV for different $U_{DC}$ (See fig. S21 for the complete data set). The solid lines are fits to the data according to the equation in the caption of fig. S21. (**F**) $T_{2,\text{Rabi}}$ time and Rabi rate $\Omega$ as a function of $U_{DC}$ extracted from the fits shown in (E). The error bars represent the standard deviation of fitted values (The error of $\Omega/2\pi$ is within the symbol size).